\documentclass[sigconf]{acmart}

\AtBeginDocument{%
  }

\setcopyright{acmlicensed}
\copyrightyear{2024}
\acmYear{2024}
\setcopyright{acmlicensed}
\acmConference[MMGR '24] {Proceedings of the 2nd International Workshop on Deep Multimodal Generation and Retrieval}{October 28, 2024}{Melbourne, VIC, Australia.}
\acmBooktitle{Proceedings of the 2nd International Workshop on Deep Multimodal Generation and Retrieval (MMGR '24), October 2, 2024, Melbourne, VIC, Australia}
\acmISBN{979-8-4007-1202-9/24/10}
\acmDOI{10.1145/XXXXXX.XXXXXX}

\usepackage{multirow}
\usepackage{algorithm}
\usepackage{algpseudocode}

\begin{document}

\title{A Learnable Agent Collaboration Network Framework for Personalized Multimodal AI Search Engine}

\author{Yunxiao Shi}
\email{Yunxiao.Shi@student.uts.edu.au}
\affiliation{%
  \institution{University of Technology Sydney}
  \city{Sydney}
  \state{NSW}
  \country{Australia}
}

\author{Min Xu}
\authornote{Corresponding author.}
\email{Min.Xu@uts.edu.au}
\affiliation{%
  \institution{University of Technology Sydney}
  \city{Sydney}
  \state{NSW}
  \country{Australia}
}

\author{Haimin Zhang}
\email{Haimin.Zhang@uts.edu.au}
\affiliation{%
  \institution{University of Technology Sydney}
  \city{Sydney}
  \state{NSW}
  \country{Australia}
}

\author{Xing Zi}
\email{Xing.Zi-1@student.uts.edu.au}
\affiliation{%
  \institution{University of Technology Sydney}
  \city{Sydney}
  \state{NSW}
  \country{Australia}
}

\author{Qiang Wu}
\email{Qiang.Wu@uts.edu.au}
\affiliation{%
  \institution{University of Technology Sydney}
  \city{Sydney}
  \state{NSW}
  \country{Australia}
}

\renewcommand{\shortauthors}{Yunxiao et al.}

\begin{abstract}
Large language models (LLMs) and retrieval-augmented generation (RAG) techniques have revolutionized traditional information access, enabling AI agent to search and summarize information on behalf of users during dynamic dialogues. Despite their potential, current AI search engines exhibit considerable room for improvement in several critical areas. These areas include the support for multimodal information, the delivery of personalized responses, the capability to logically answer complex questions, and the facilitation of more flexible interactions. This paper proposes a novel AI Search Engine framework called the Agent Collaboration Network (ACN). The ACN framework consists of multiple specialized agents working collaboratively, each with distinct roles such as Account Manager, Solution Strategist, Information Manager, and Content Creator. This framework integrates mechanisms for picture content understanding, user profile tracking, and online evolution, enhancing the AI search engine's response quality, personalization, and interactivity. A highlight of the ACN is the introduction of a Reflective Forward Optimization method (RFO), which supports the online synergistic adjustment among agents. This feature endows the ACN with online learning capabilities, ensuring that the system has strong interactive flexibility and can promptly adapt to user feedback. This learning method may also serve as an optimization approach for agent-based systems, potentially influencing other domains of agent applications.
\end{abstract}




\maketitle

\section{Introduction}
In today's information-saturated world, information retrieval systems play a crucial role in sifting through vast data to find content that resonates with individual needs, thereby alleviating the problem of information overload. For years, traditional search engines like Google and Bing have been the primary tools for this task. However, recent advancements in large language models (LLMs) \cite{zero_shot_llm, PaLM} and retrieval-augmented generation (RAG) techniques \cite{improve_by_retrieve, RAG, four_module} have given rise to a new generation of AI-powered search engines, such as Perplexity and Tiangong. These innovations have revolutionized information access by shifting from static query inputs to interactive dialogues with AI agents. Instead of manually browsing through multiple web pages, users can now rely on AI agents to synthesize and present the most relevant information to meet the information gaining requirements.

However, current AI search engines still have several aspects that need improvement. \textit{(1) Multimodal Information Support}. Existing AI search engines primarily generate pure text content, whereas web content encompasses various modalities including text, images, tables, and videos \cite{ji2024MMGR}. The support of multimodal content understanding is essential for yielding high response quality and rich presentation content. \textit{(2) Personalized Response}. Current AI search engines deliver uniform content to different users, overlooking the key factor of personalization and customization. While traditional search engines have incorporated some personalization features \cite{Encoding_history,how_whether,Explicit_Subtopics}, AI search engines have yet to effectively integrate this aspect. For instance, when I asked GPT-4, Perplexity for muscle-building diet recommendations as an Indian, they all suggested beef as a primary protein source, which contradicts the cultural and dietary restrictions of Indians. \textit{(3) Answering Complex Logic Requirement}. Current AI search engines can handle simple information retrieval and generation tasks but struggle with complex, logic-intensive queries. Such queries often require multi-keyword searches and iterative retrieval processes, and the generated information needs logical coherence and strategic formulation. \textit{(4) Timely Learning and Adjustment}. Current AI agents are "expert-centric", relying heavily on pre-set prompts and workflows \cite{zhou2024symboliclearningenablesselfevolving}, restricting their ability to autonomously adapt based on users' feedback. 

Motivated by the aim of addressing these limitations, we propose a AI search engine framework named the Agent Collaboration Network (ACN). This framework comprises multiple agents, each performing distinct roles, including Account Manager, Solution Strategist, Information Manager, and Content Creator. The Account Manager interacts with users, tracks user profiles, gathers feedback, and transfer user information searching requirements to the Solution Strategist. The Solution Strategist takes account the user profile, uses a chain of thought method to solve complex user requirement step-by-step, and logically plans the article outline. It allocates information retrieval tasks to the Information Manager and content generation tasks to the Content Creator. Once everything is ready, the Solution Strategist triggers the Finalize Article action, completing the generation of multimodal content and delivering the results to the Account Manager. The Information Manager handles multimodal information retrieval, while the Content Creator generates multimodal content tailored to specific users based on the Solution Strategist's instruction and user profiles.

Additionally, we have designed an optimization algorithm named Reflective Forward Optimization (RFO) for the ACN, which can automatically adjust based on user feedback. We first design a LLM-based optimizer, which can inspect intermediate results within the agent-to-agent workflow and generates reflective reviews based on a given feedback. These reviews help improve adjustable parameters such as agent prompts, function parameters, and system settings while providing further feedback to the called agent. By running the RFO algorithm along the response-generating agent call stack, we obtain a collection of reviews for each agent. We then aggregate all review suggestions and use the LLM to update each agent, ultimately refining the entire ACN. This timely online-learning method enhances the flexibility of interactions and aligns the ACN more closely with the user's requirement.

In summary, our work provides several key contributions:
\vspace{-7mm}
\begin{itemize}
  \item We propose a novel AI search engine framework named Agent Collaboration Network (ACN), which incorporates a specially designed agent-learning method called Reflective Forward Optimization (RFO). The ACN surpasses traditional AI search engines by supporting multimodal content output, personalized content generation, and the creation of more logically structured and complex information. Additionally, it can continuously adjust and learn based on user feedback promptly.
    \item We design a synthetic dataset and use LLM-played judger to verify the effectiveness of ACN compared to SOTA TianGong and Perplexity AI search engines, demonstrating its superior ability to generate engaging information with multi-modality, logical-well, useful content, provide personalized user experience. 
    \item We point out the current research gap in evaluating AI search engines' responsiveness to user feedback. We have analyzed the feasibility of experiments and outlined future plans to address this deficiency.
\end{itemize}

\section{Related Works}
\subsection{AI Search Engine}
AI search engines represent the convergence of large language models (LLMs), retrieval-augmented generation (RAG), and intelligent agent technologies, heralding a new era of search engine innovation. Given the nascent stage of AI search engine technology, we categorize the information retrieval process into six distinct phases \cite{modular_rag_survey}: identification of information retrieval requirements  \cite{TOC}, retrieval augmentation \cite{Query_Rewriting}, information retrieval and knowledge gathering \cite{assessing_adaptive_retrieval}, knowledge caching \cite{Expel}, knowledge filtering and ranking  \cite{NLI, LLM_filter}, and LLM-based content generation, followed by verification and refinement \cite{RARR}. Certain AI search engines do not adopt a conversational interface, requiring users to input search queries into a search bar, thereby omitting the initial phase of identifying information retrieval needs. Retrieval enhancement can be selectively integrated before the information retrieval and knowledge gathering phase, generating multiple search keywords to ensure a higher recall rate of relevant knowledge. Knowledge caching post-retrieval mitigates resource consumption associated with the retrieval and gathering phase, enhancing system responsiveness and efficiency. Knowledge filtering and ranking technique is employed before the generation phase, eliminate irrelevant information, thereby enhancing the precision of the retrieved knowledge, improving the robustness of LLM responses, and ensuring content quality. The verification and refinement phase post-generation ensures the factual accuracy of the AI-generated content and optimizes its presentation format for the user. Therefore, among these six phases, the steps of information retrieval and knowledge gathering, as well as LLM-based content generation, are necessary, corresponding to the retrieve-then-read pipeline of RAG. The other steps are optional according to practical application scenarios.

The advanced studies mentioned above focus primarily on ensuring the precision and efficiency of information generation in AI search engines. Beyond these critical metrics, our study more emphasizes content richness, personalization, and interactivity. These factors are crucial for maintaining the attractiveness of the content and enhancing the overall user experience with AI search engines.

\subsection{Personalized Generation of LLM}

Recent studies \cite{kirk} emphasize the importance of personalizing large language models (LLMs) beyond aggregate fine-tuning methods like RLHF, as these may not fully capture diverse user preferences and values. Micro-level preference learning can better align models with individual users. Current personalization techniques mainly involve prompt tuning, which models user profiles based on historical search data to prompt LLM generating tailored outputs. Basic approaches use the entire user action history for prompting, while more advanced methods selectively retrieve relevant user data using memory mechanisms \cite{CoPS}. To address potential information loss, \cite{Richardson2023Integrating} proposes a task-aware user profile summarization for prompting. Another approach \cite{baek2024knowledge} constructs knowledge graphs from user search and browsing activities to enhance prompt relevance.

While existing work focuses on prompt construction and user profiling, our Agent Collaboration Network (ACN) architecture shifts the emphasis to tuning agents across each step of the AI search engine workflow, aiming for a more integrated and efficient personalization strategy.

\begin{figure*}[!t]
  \centering
    \includegraphics[width=6.3 in]{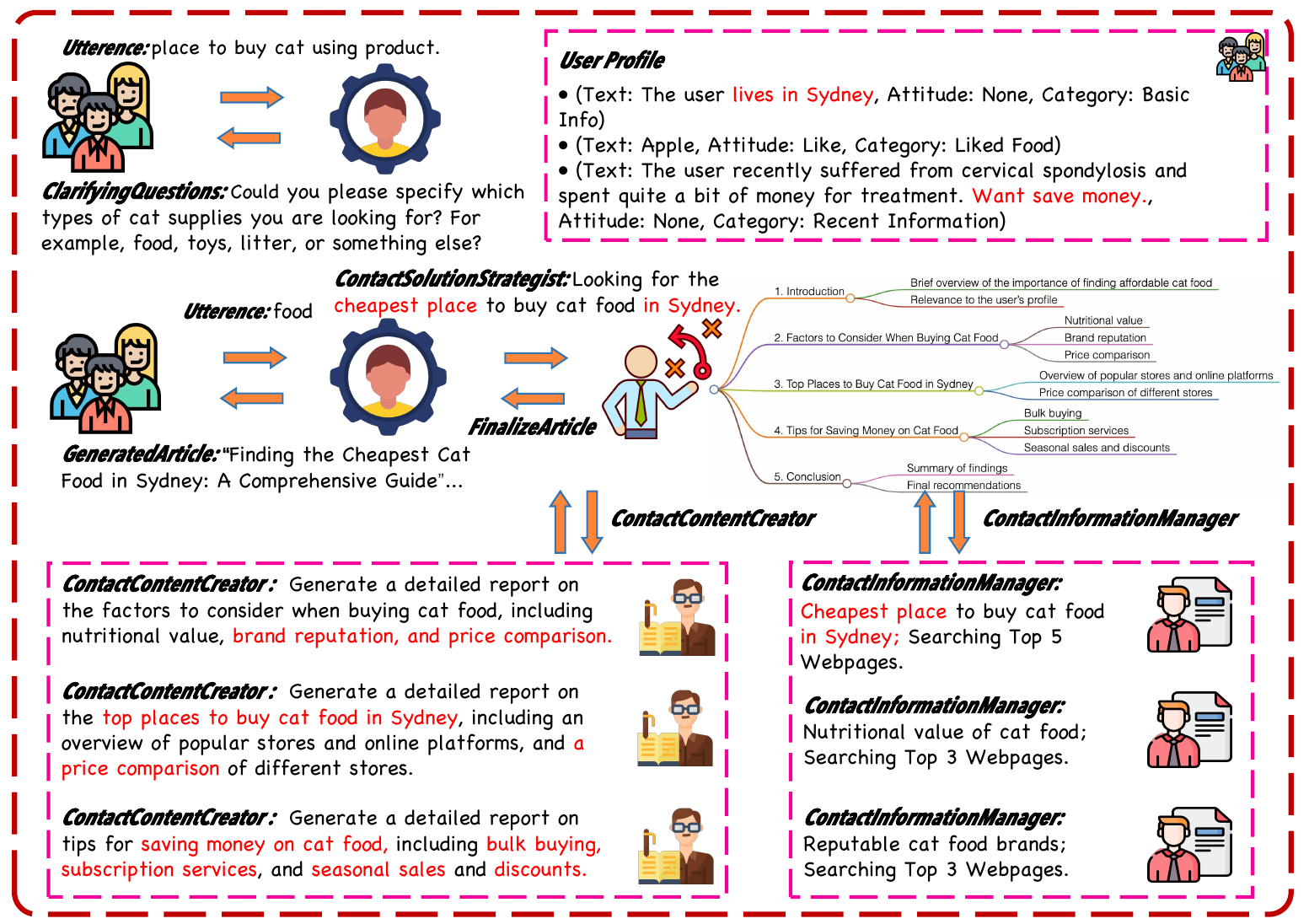}
    \vspace{-2mm}
   \caption{Agent Collaboration Network Framework with a case study "Place to buy cat using product." The red text illustrates how the ACN customizes message passing, the information searching process, and the information generation process to the specific user. The mind map is generated using COT, guiding the function calling for solving the user's information gaining requirements in a logic, deep, and structural way.}
    \label{fig: ACN_Framework}
\end{figure*}

\subsection{AI Agent}
AI agents can interact with environments, dynamically selecting optimal actions to achieve predetermined objectives. With the advent of LLMs capable of function calls \cite{Gpt4tools,Toolllm}, AI agents driven by such models exhibit exceptional intelligence and flexibility. This advancement paves the way for the evolution of RAG technology towards an Agentic RAG paradigm\footnote{https://github.com/infiniflow/ragflow}. Recent studies have introduced frameworks where multiple interconnected agents collaborate to support a wide array of tasks \cite{MMedAgent,gao2024simulatingfinancialmarketlarge}. Research indicates that agents can enhance their capabilities through reflective thinking \cite{cheng2024lift} or by leveraging optimization techniques analogous to neural networks \cite{zhou2024symboliclearningenablesselfevolving}, highlighting significant potential for online, training-free self-adjustment of Agent-based applications.

Our work pioneers a universal framework for AI search engines utilizing multi-agent collaboration named ACN. We also introduce an optimization method enabling real-time learning and adaptation based on user feedback. Such adaptability enhances the personalized and interactive capabilities of AI search engines, offering a more tailored and responsive user experience.

\section{Agent Collaboration Network}
The proposed ACN framework comprises multiple agents, include the Account Manager, Solution Strategist, Information Manager, and Content Creator. Each one with distinct roles and responsibilities, and they collaborate dynamically to deliver satisfying response. The framework with a case study is presented in Figure~\ref{fig: ACN_Framework}.

\subsection{Account Manager}
The Account Manager Agent plays a pivotal role in engaging with users to comprehend their needs, monitor their interests, and assist them in articulating precise requirements. The Account Manager also serves as a critical communication conduit to Solution Strategist agent. When confirming the users' specific information retrieval needs, it conveys detailed user requirement to the Solution Strategist agent. 

The Account Manager continuously tracks users' interest and information for building their profile. For a given user $ u $, let $ P = \{ d_1, d_2, \ldots, d_{|P|} \} $ represent the user profile, where each element $ d_i = (\text{text}, \text{attitude}) \in P $ encapsulates a concise description of the user's fundamental information or interest preferences. For  $ d_i $ detailing basic information, the attitude is labeled as 'None', whereas for interest preferences, the attitude is categorized into $\{\text{Positive}, \text{Neutral}, \text{Negative}\}$. We use $ d_{\text{new}} $ to denote the captured profile description during the Account Manager's ongoing interactions with the user. To integrate this new description into the pre-existing set $ P $, we assess the topic similarity between $ d_{\text{new}} $ and each $ d_i $, represented as $ S(d_{\text{new}}, d_i) $. This assessment leverages the bge-m3 model, which generates both dense and sparse embeddings—also referred to as lexical weights—for the text of each description. This dual embedding approach enables a hybrid similarity computation, effectively balancing keyword matching with semantic alignment between descriptions. We establish a similarity threshold $\gamma$; if $S(d_{\text{new}}, d_i) \geq \gamma $, $ d_{\text{new}} $ replaces $ d_i $, otherwise, $ d_{\text{new}} $ is appended to the set $ P $ as a new element.

The user's feedback can trigger the Account Manager Agent's function of Accepting Feedback and Reflection, the feedback will then be conveyed to the ACN optimizer for improving the collaborations of ACN. The details are described in Section~\ref{sec: RFO}.

\subsection{Solution Strategist}
The Solution Strategist Agent adhere to the detailed requirements from Account Manager Agent, and take account users' profile, then meticulously plans and orchestrates the process of addressing the information retrieval and generation task. By leveraging LLMs and employing the chain of thought method, the agent generates a structured and detailed pathway for problem-solving. As part of its strategic plan, the Solution Strategist Agent is empowered to execute specific actions such as Search Information, Generate Content, and Finalize Article.

For the action Search Information, the Solution Strategist Agent allocates a retrieval task to the Information Manager Agent, providing a precise search query. In the case of Generate Content, a text generation task is assigned to the Content Creator Agent, accompanied by a detailed creation requirement. Finally, for the action Finalize Article, the Solution Strategist Agent merges the generated content and delivers it back to the Account Manager Agent.

\subsection{Information Manager}
The Information Manager Agent is dedicated to retrieving pertinent information, utilizing the Bing Search Engine v7 API to access real-time and up-to-date web content. This agent retrieves and converts webpage content into markdown format, ensuring the inclusion of all text and image links.

Given the web's vast repository of data, much of it can be irrelevant, potentially obscuring the LLMs' comprehension of key information. This irrelevance can degrade the quality of generated content and lead to unnecessary token consumption. Therefore, it is crucial to filter out non-essential content. Initially, the webpage content is divided into segments based on double newline characters, with each segment treated as an independent chunk. The bge-m3 model is then employed to calculate the similarity score of each chunk relative to the query. A similarity threshold $\lambda$ is established, and chunks with similarity scores below $\lambda$ are filtered out, ensuring that only the most relevant information is retained.

To understand the contextual and semantic information of images embedded within the text, we select the contextual content surrounding each <img> tag, capturing the relevant text that provides insight into the image's role within the document. We then task vision language models (VLMs) with inferring the caption of the image and describing its content in a concise manner. This process yields two outputs: a descriptive caption and a succinct content summary. These outputs, along with the image's URL, form a comprehensive set of image-related data. This data is archived for later usage of content generation tasks.

\subsection{Content Creator}
The Content Creator Agent is responsible for adhering to the creation requirements specified by the Solution Strategist Agent, generating personalized reports with multimodality. This process is meticulously designed to align with the user's profile, ensuring a high degree of personalization and relevance in the generated content.

The LLMs' generation is prompted with user's profile, retrieved external knowledge, and the image information. The profile prompt the the generated content aligned with the user's interests, preferences, and needs, ensuring that the generated article is not only informative but also appealing and useful to the user. During the generation, the agent can generate image captions, allowing for the collected image to be integrated smoothly into the report.

\subsection{Role Setting}
In designing the Account Manager Agent and Solution Strategist, we leveraged the function calling capabilities of LLMs to achieve a higher degree of flexibility. For the Content Creator Agent, which is tasked solely with text generation, we opted to utilize the traditional text completion ability of LLMs.
Detailed function calling settings are available in Appendix~\ref{sec: functionc_call_prompt}

\subsubsection{Prompt for Account Manager}

\quad

\noindent \textbf{Instruction}: You are Account Manager in a collaborative agent network aims at providing Personalized Multimodal Information Retrieval and Generation service. Your task is to interact with users in a friendly manner, maintain relationships with customers, ensure customer satisfaction, and understand their needs and expectations through ongoing communication. Furthermore, you are responsible for coordinating the company's Solution Strategist Agent for solving customers' personalized multimedia information retrieval and generation request.

\noindent \textbf{Functions}: Normal Reply, Clarifying Questions, Providing Suggestions, Contact Solution Strategist, Tracking User Preferences, Accepting Feedback and Reflection.

\subsubsection{Prompt for Solution Strategist}

\quad

\noindent \textbf{Instruction:} You are Solution Strategist in a collaborative agent network aims at providing Personalized Multimodal Information Retrieval and Generation service. Your task is to develop a logical plan to solve the tasks described in the \textit{[User Requirement]} conveyed from the Account Manager agent. You should outline this plan step by step, using flexible combinations of calling Search Information and Generate Content. But you must end with the function Finalize Article. Besides, you should also consider the provided \textit{[User Profile]} to make your logical plan specialized for the user.

\noindent \textbf{User Requirement:} \textit{\{User Requirement is here.\}}

\noindent \textbf{User Profile:} \textit{\{User Profile is here.\}}

\noindent \textbf{Functions:} Search Information, Generate Content, Finalize Article.

\subsubsection{Prompt for Content Creator}

\quad

\noindent \textbf{Instruction}: You are Content Creator Agent in a collaborative agent network aims at providing Personalized Multimodal Information Retrieval and Generation service. Your task is to utilize your professional content creation skills, based on the provided \textit{[External Knowledge]}, and \textit{[Image Source]} as your reading material to generate a detailed multimodal content in markdown format. You should strictly follow the \textit{[Writing Requirement]}, and include appropriate images as much as possible to make the content rich. You also must consider the \textit{[User Profile]}, and makes the content personalized, aligning with user's information, preferences.

\noindent \textbf{External Knowledge}: \textit{External Knowledge is here.}

\noindent \textbf{Image Source}: \textit{Image Source is here.}

\noindent \textbf{Writing Requirement}: \textit{Writing Requirement is here.}

\noindent \textbf{User Profile}: \textit{User Profile is here.}

\begin{figure}[!t]
  \centering
    \includegraphics[width=3.3 in]{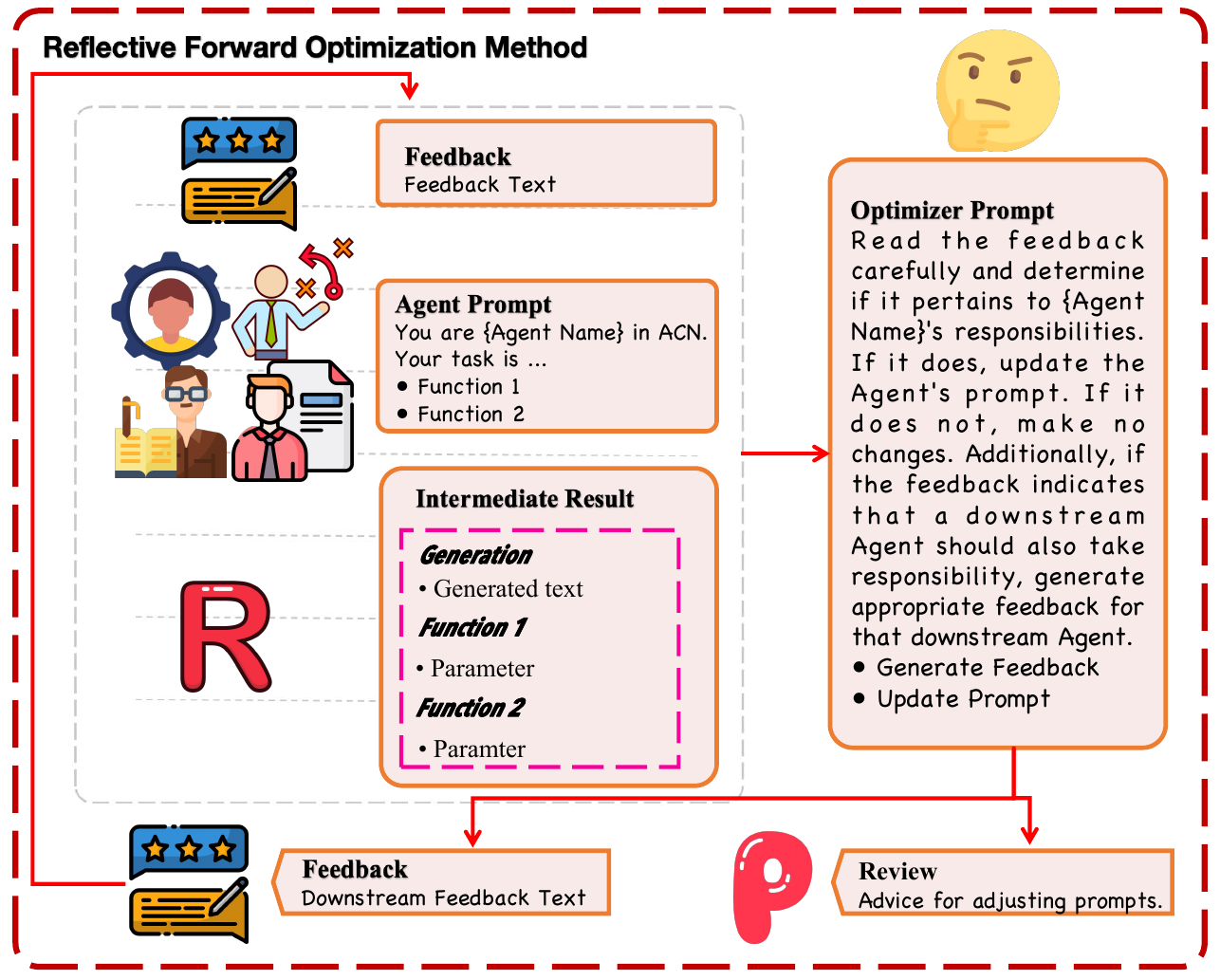}
    \vspace{-3mm}
   \caption{RFO algorithm workflow.}
    \label{fig: RFO}
\end{figure}

{
\setlength{\textfloatsep}{15pt}  

\begin{algorithm}
\caption{Reflective Forward Optimization (RFO)} \label{alg: RFO}
\begin{algorithmic}[1]
\State \textbf{Input:} User Feedback $UFB$, Agent Collaboration Network $ACN$, ACN Intermediate Result $RESULT$
\State \textbf{Output:} Optimized Agent Collaboration Network $OACN$

\Function{RFO}{$UFB, ACN, RESULT$}
    \State Initialize $FB \gets UFB$
    \State Initialize $stack \gets [(\mathcal{A}, UFB)]$ \Comment{Initialize stack with root agent and user feedback}
    
    \While{$stack$ is not empty}
        \State $(A, FB) \gets \text{stack.pop()}$ \Comment{Current Agent $A$, Current Feedback $FB$}
        \State $prompt \gets A.prompt$
        \State $result \gets RESULT.A$

        \State $(Down\_Agents, Down\_FBs, Prompt\_Review) \gets \text{Optimizer}(FB, prompt, result)$
        
        \State $A.Review\_List \gets Prompt\_Review$
        
        \For{each $(Down\_Agent, Down\_FB)$ in zip($Down\_Agents, Down\_FBs$)}
            \State $\text{stack.push}(Down\_Agent, Down\_FB)$
        \EndFor
    \EndWhile

    \For{each $Agent$ in $ACN$}
        \State $Agent.prompt \gets \text{UpdatePrompt}(Agent.Review\_List)$
    \EndFor

    \State \Return $OACN$
\EndFunction

\end{algorithmic}
\end{algorithm}

\section{Reflective Forward Optimization}
\label{sec: RFO}
In the process of delivering services to users, they may provide feedback and the Account Manager agent can automatically utilizes to trigger the function of Accepting Feedback and Reflection. This function is essential for the adaptive optimization of the Agent Collaboration Network, enabling real-time, conversational online adjustments. We have developed a novel optimization method termed Reflective Forward Optimization (RFO) to enhance the agent network.

The adjustable parameters for each agent in the ACN are all prompts. Unlike the backpropagation optimization algorithm in neural networks, our RFO is based on a depth-first traversal algorithm for forward propagation optimization. Using an LLM-based optimizer, it systematically reflects on and examines the previous processes used to fulfill user requirements. This allows the algorithm to assign responsibility to agents and make necessary adjustments. The illustrative workflow of RFO is in Figure~\ref{fig: RFO}, and the detailed algorithmic process is in Algorithm~\ref{alg: RFO}. 

The design of optimizer is as follows:
\noindent \textbf{Instruction}: You are an optimizer based on a large language model. The task involves:
(1) There is a [Call Agent] that passes parameters [Message] to a [Called Agent]. The external input to the [Call Agent] is [Input], and the output from the [Call Agent] to the external environment is [Output]. (2) The [Call Agent] can adjust the parameters in [Parameter]. (3) The external environment provides [Feedback] on [Output]. Your task is (a) Determine if the cause of the [Feedback] lies with the [Call Agent]. If it does, you need to review each parameter in [Parameter] one by one and provide adjustment suggestions in <review>. (b) If the cause is not with the [Call Agent], you need to provide downstream feedback to the [Called Agent] in the <down\_feedback>, to let the [Called Agent] to further reflect himself. If the [Called Agent] is None, then just set <down\_feedback> as None. We must use function: Optimize to generate the <review> and <down\_feedback>.

\noindent \textbf{Functions}: Optimize.

}

\section{Experimental Setup}
\subsection{Dataset}
Personalized dialogue datasets, such as CONVAI2 \cite{ConvAI2}, DuLeMon \cite{xu2022long}, and KBP \cite{KBP}, emphasize enhancing the personalization of conversations. These datasets typically include a segment of the user profile, ensuring that the generated dialogues align with this profile. The LaMP \cite{LaMP} benchmark is specifically designed to train and evaluate large language models (LLMs) for personalized outputs, with tasks ranging from personalized title generation to tweet paraphrasing in LaMP4-9. Although these datasets are relevant to our research, they do not fully address our specific focus. Our objective is to design a dataset comprising multiple session chat records, each containing varied-length dialogues between users and AI search engines. Users demonstrate dynamic topic interests, provide feedback, and articulate multimodal information requirements. In response, AI search engines not only engage in basic chat but also generate informative responses.

In light of the lack of appropriate datasets and inspired by recent advancements in utilizing the role-playing capabilities of large language models (LLMs) for simulating realistic scenarios to generate datasets\cite{ abdullin2024syntheticdialoguedatasetgeneration, let_llms_talk}, we introduce the synthetic Multi-Session Multi-Turn Personalized Information Inquiry and Generation (MSMTPInfo) dataset. This dataset is meticulously designed to emulate interactions between authentic users, characterized by distinct and evolving personalities, and an AI search engine. Each session comprises multiple turns of dialog between the user and the AI search engine, forming a series of user utterances and corresponding responses.

The dataset spans conversations across 13 diverse main themes and many sub-themes. Initially, we delineate the primary and secondary themes within the conversational content and elucidate the specific actions associated with the attitudes exhibited in user responses, as illustrated in Figure~\ref{fig: topic_attitude}. Following this, we utilize the prompt template shown in Figure~\ref{fig: Prompt_MSMTPInfo} to systematically generate data between a user and an AI assistant on a session-by-session basis.

\begin{figure*}[!t] 
  \centering
    \includegraphics[width=6.3 in]{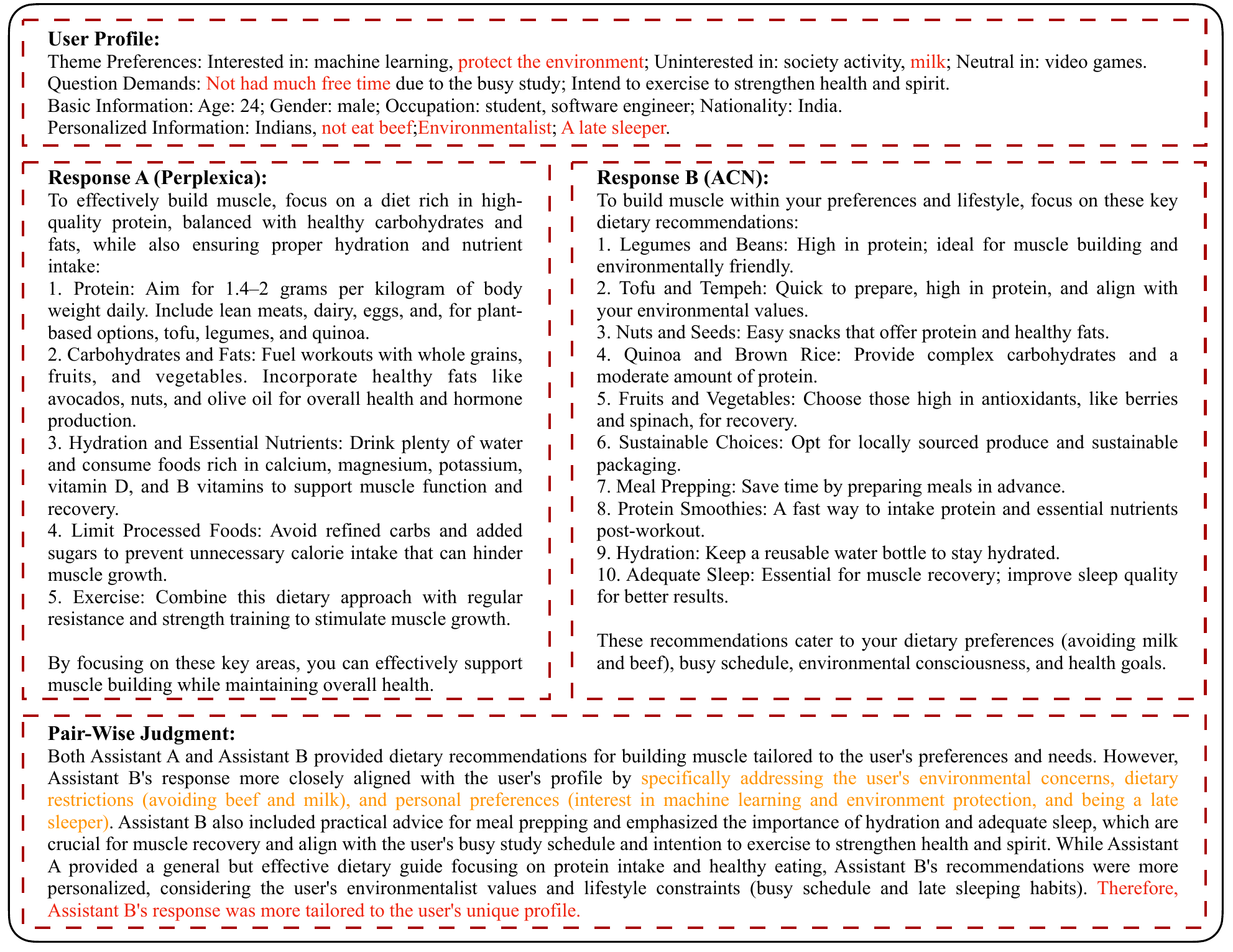}
    \vspace{-3mm}
   \caption{Comparison of AI Search Engine Responses to the Query "Give me a dietary recommendation for building muscle."  A LLM played judge subsequently determines that Response B (ACN) is better.}
    \label{fig: case_study}
\end{figure*}

\begin{figure*}[!h] 
  \centering
    \includegraphics[width=6.3 in]{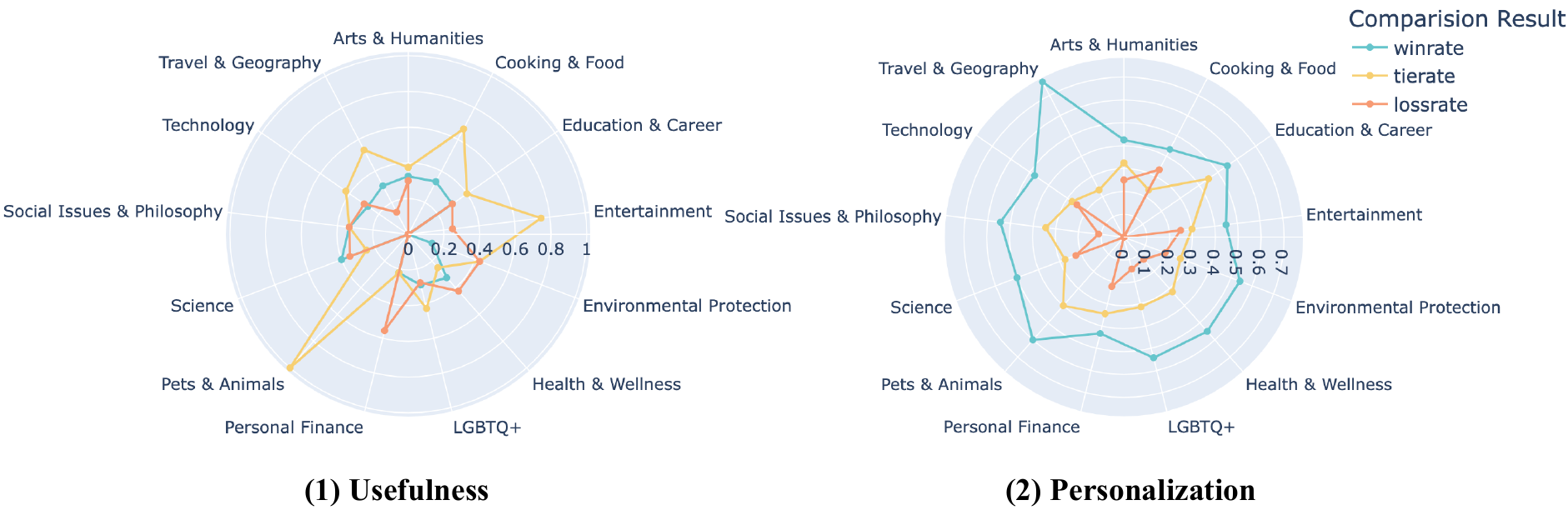}
    \vspace{-3mm}
   \caption{The results of pairwise comparisons between Basic and ACN responses across all categories on the MSMTPInfo dataset.}
    \label{fig: pair_result}
\end{figure*}

\subsection{Baselines and Our Method.} 
In this study, we evaluate the performance of several notable commercial AI search engines. Besides, given the impracticality of conducting fair comparisons due to discrepancies in the number of indexed webpages, variations in backbone LLM models, and the inability to modify these closed-source commercial engines for rigrous studies, we also consider one open-source AI search engine. 

1. Perplexity\footnote{https://www.perplexity.ai/}: Widely-used and popular all over the world.

2. TianGong\footnote{https://www.tiangong.cn/}: It has research mode capabilities that produce meticulously logical and comprehensive reports.

3. Perplexica\footnote{https://github.com/ItzCrazyKns/Perplexica/tree/v1.3.0}: It is can be considered as open-source counterpart of Perplexity. We set each query searching 5 webpages via Bing, utilizing GPT-4o-mini as the backbone LLM.

4. Our ACN: we integrated our proposed ACN framework into the open-soure Perplexica framework, supporting multimodal text and image outputs, personalized information generation, logical planning for complex queries, and adaptive online adjustments.

\subsection{Evaluation Setup.} 
Our evaluation methodology leverages insights from prior research on using LLM-as-a-judge \cite{MTbench} method, which demonstrated that employing GPT-4 as an adjudicator aligns with human assessments at an agreement rate exceeding 80\%. In our approach, we present the LLM-based judger with two responses: one generated by the ACN and the other from alternative configurations. The judger is tasked with distinguishing between these responses and delivering a judgment based on predefined criteria of win, loss, or tie. To mitigate any positional bias inherent in LLMs, we subsequently reverse the positions of the two responses within the prompt and re-evaluate. If the judger's decisions are consistent across both evaluations—whether both are wins, ties, or losses—the final judgment is accordingly recorded as a win, tie, or loss. Conversely, if the adjudications differ, the final result is deemed a tie. Finally, we calculate the win rate, tie rate, loss rate of ACN compared to other alternative configuration. Our evaluation encompasses multiple dimensions to provide a comprehensive assessment:

$\bullet$ \textbf{Content Richness}: The richness of the content, including both textual and visual elements, is vital for capturing and sustaining user interest.

$\bullet$ \textbf{Information Usefulness}: Despite the richness of content, the actual utility of the information is paramount. An abundance of redundant or irrelevant information fails to meet the user's need for efficient knowledge acquisition. Hence, the AI search engine's effectiveness should be further evaluated from the perspective of information usefulness.


\begin{figure*}[!t]
  \centering
    \includegraphics[width=6.3 in]{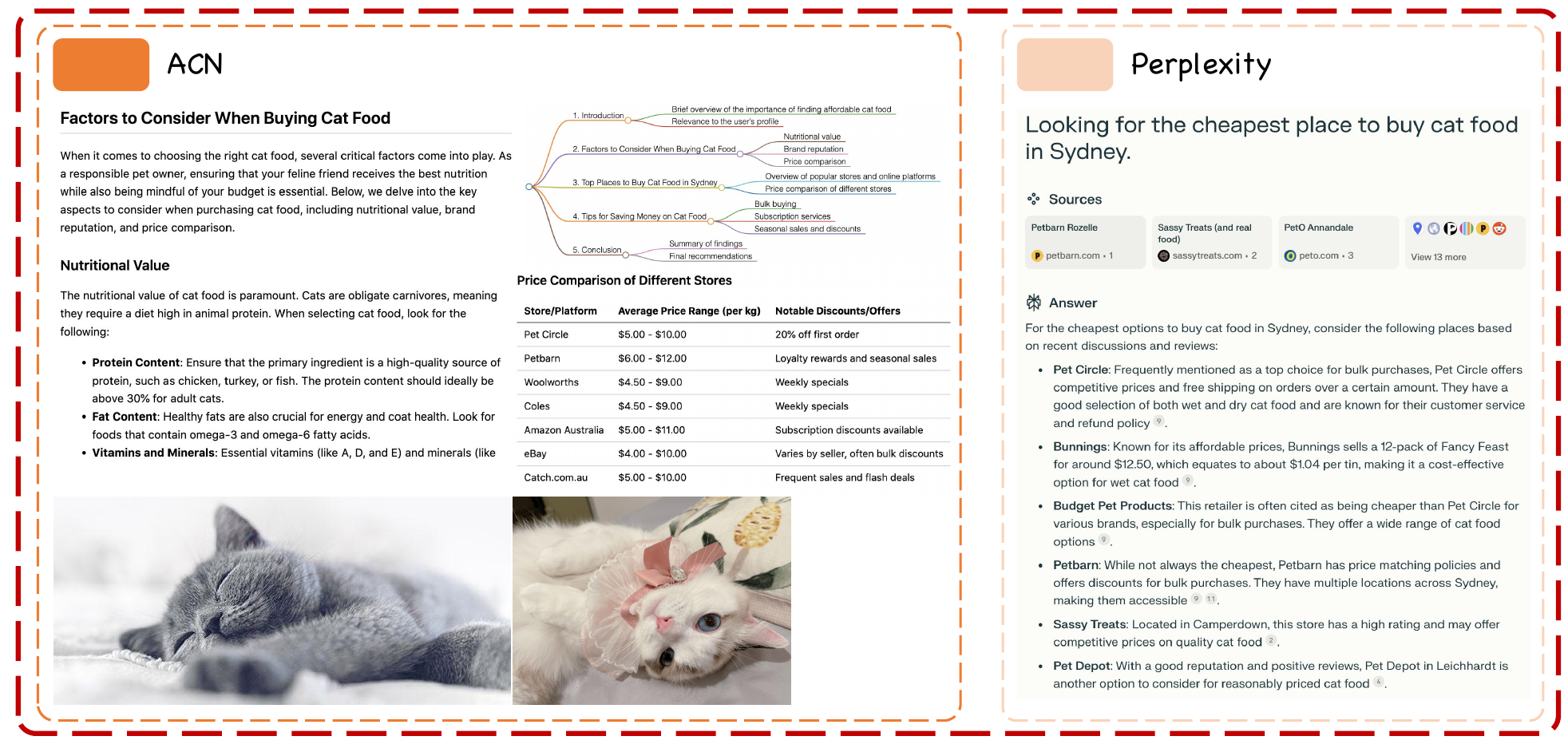}
    \vspace{-5mm}
   \caption{Comparison between ACN and Perplexity.}
    \label{fig: rishness}
\end{figure*}

$\bullet$ \textbf{Content Personalization}: Leveraging user profile information revealed either in prior sessions or during the ongoing interaction is essential for customizing the dialogue. This personalized approach is a key factor in delivering a superior user experience through the AI search engine.

$\bullet$ \textbf{Writing Logicality}: AI search engines can address complex queries that traditional search engines fail to resolve. These questions often require more than simple keyword searches, as they encompass intricate problems with inherent logical structures.

\section{Experimental Results}
We now present our part of experimental results, and report findings from various auxiliary studies and analyses.

\subsection{Richness Analysis}
Our proposed ACN demonstrates a substantial superiority over current AI search engines that are limited to text-based article generation. By incorporating multiple modalities, including text, appropriate image insertions, and tabular data presentation, ACN-generated articles offer enhanced visual appeal and engagement. This multimodal approach not only captivates readers' attention but also provides a better satisfying experience. Figure~\ref{fig: richness} exemplifies this advantage.

\subsection{Usefullness and Personalization Analysis}
The whole dataset's comparative assessment outcomes in different topics are statistically summarized and visualized via a radar chart in Figure~\ref{fig: pair_result}. The left subplot of the figure exclusively concentrates on criteria usefulness. Subsequently, right subplot of the figure evaluates personalization by assessing how well the responses aligned with the user's profile. It was observed that response A and response B often reached a draw across various topics, with instances of unilateral victories or defeats being comparatively rare. However, when takinging into personalization, response B consistently outperformed response A, demonstrating a superior ability of ACN delivering tailored responses that align more closely with the user's unique profile.

Further, a case study of the pair-wise judgement result is illustrated in Figure~\ref{fig: case_study}. This case study provides a brief summary of two AI search engines' responses to the identical user query. The response generated by Perplexica (anonymized as Response A) does not take the user profile into account. In contrast, the response generated by the ACN (anonymized as Response B) consistently tracks the user profile and produces a more customized recommendation. This investigation elucidates the comparative analysis of response between two AI search engines. The focal point of this case study hinges on the assessment of responses provided by A and B in the context of their consideration of the user profile. The judgement result unveils that while both responses delivered accurate and comprehensive response. But response B is distinguished by the personalized nature, tailored specifically to the user's preferences and needs.

\vspace{-7pt}
\subsection{Logicality Analysis}

We have developed a Solution Strategist Agent that leverages Chain of Thought (COT) reasoning to enhance the logical capabilities of search engines in generating responses to complex questions. To rigorously evaluate the logical soundness of ACN, we assess the following three dimensions:

$\bullet$ \textbf{Depth}: This metric evaluates the thoroughness of the strategic plan in exploring a particular point of consideration.

$\bullet$ \textbf{Comprehensiveness}: This metric measures the extent to which the strategic plan addresses both explicit and implicit factors necessary for a robust response.

$\bullet$ \textbf{Reasonability}: This metric assesses the relative rationality of the strategic plan when analogized to human problem-solving and planning processes.

\begin{figure}[!t]
  \centering
    \includegraphics[width=3.4 in]{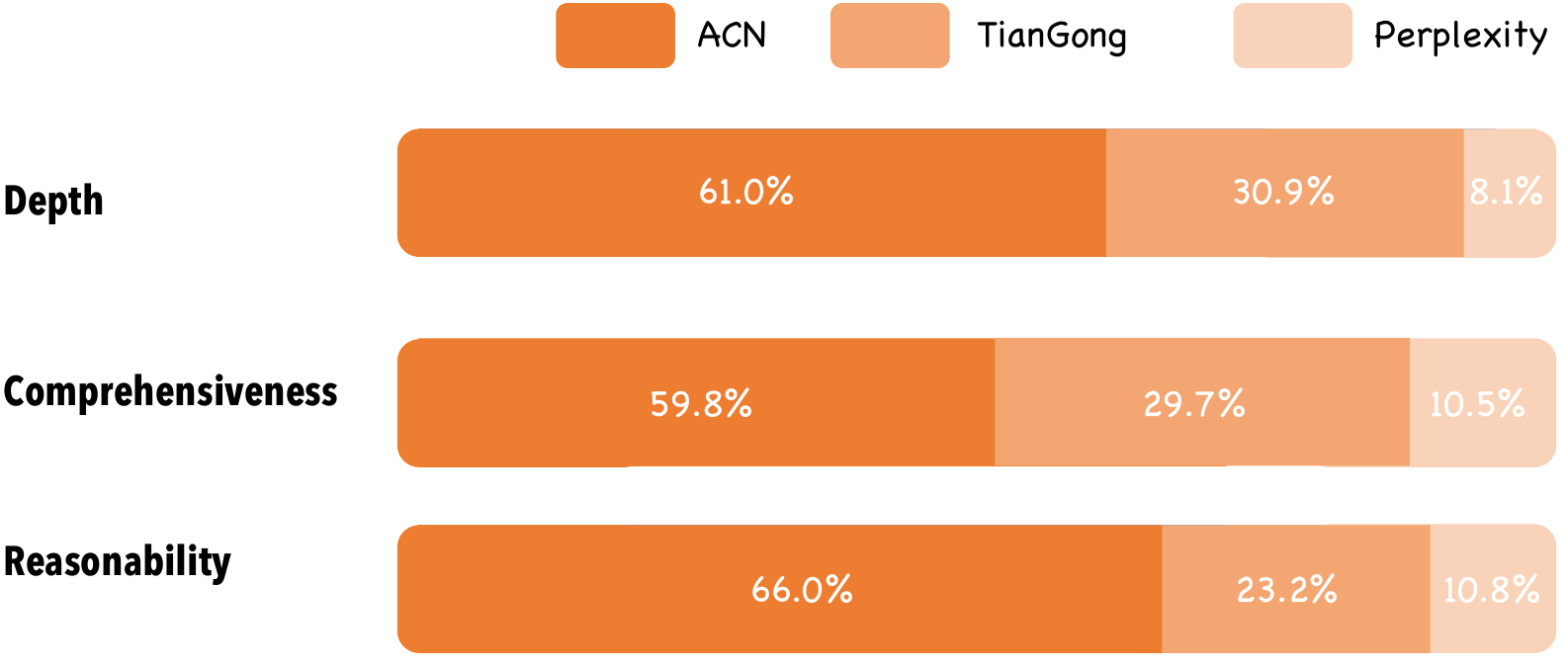}
    \vspace{-5mm}
   \caption{Comparative Evaluation Results of Pairwise Comparisons among ACN, TianGong, and Perplex. We conducted pairwise comparisons among these search engines, calculating the adjusted win rate for each and subsequently normalizing the results.}
    \label{fig: plan_compare}
\end{figure}

We benchmarked our ACN against TianGong and Perplexity. The pair-wise comparison results are illustrated in Figure~\ref{fig: plan_compare}, demonstrate a significant absolute improvement, underscoring the critical role of the Solution Strategist Agent and affirming the superiority of our proposed ACN.

\section{Discussion and Future Works}
Our current research has yet to undergo rigorous empirical validation, particularly concerning the ACN's capability of online learning and prompt adjusting based on user feedback. This limitation arises from the necessity for real human evaluations, which require more time and meticulously designed experimental protocols. To address this gap, our future research will focus on the following:

1. Identifying suitable volunteers for testing the ACN.

2. Standardizing experimental procedures by designing feedback types. Users will be able to provide feedback within predefined categories.

3. Comparing the ACN's performance with other models such as TianGong and Perplexity. Given that large models possess context-aware prompt learning capabilities, they theoretically offer some degree of real-time adjustment. However, the extent of this capability remains unclear and necessitates empirical investigation.

4. Developing metrics to evaluate the AI search engine's responsiveness to user feedback. We suppose that feedback will influence dialogue consistency and content personalization.





\begin{acks}
This work was sponsored by the \texttt{Australian Research Council under the Linkage Projects Grant LP210100129}, and by the program of \texttt{China Scholarships Council (No. 202308200014)}.
\end{acks}

\bibliographystyle{ACM-Reference-Format}
\balance
\bibliography{acmart}

\appendix

\begin{figure*}[t] 
  \centering
    \includegraphics[width=5.5 in]{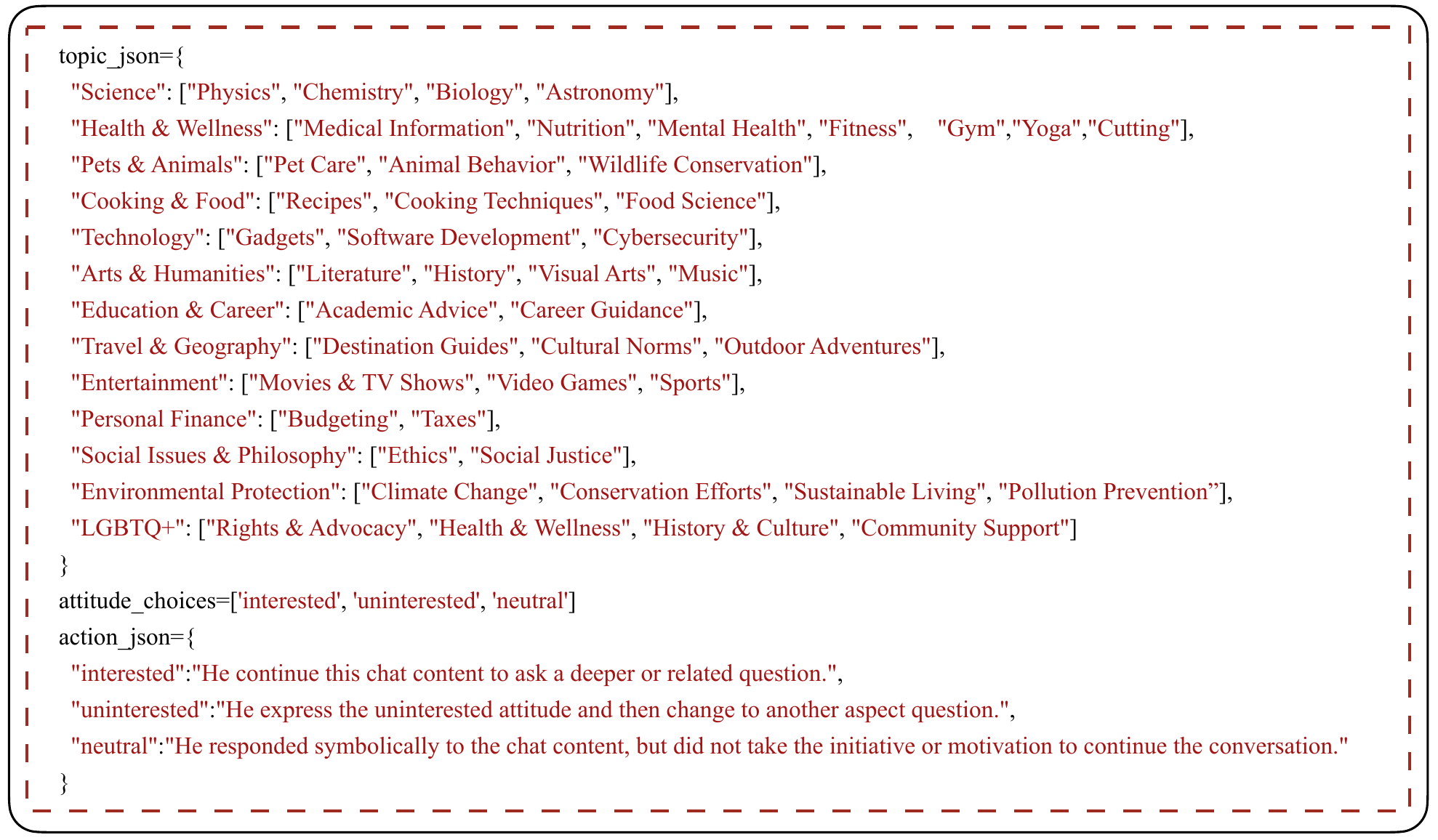}
    \vspace{-5mm}
   \caption{Thirteen main topics and subtopics of the conversational content. As well as the three types of attitude of user responses, alongside with their detailed response actions.}
    \label{fig: topic_attitude}
\end{figure*}

\begin{figure*}[t] 
  \centering
    \includegraphics[width=5.5 in]{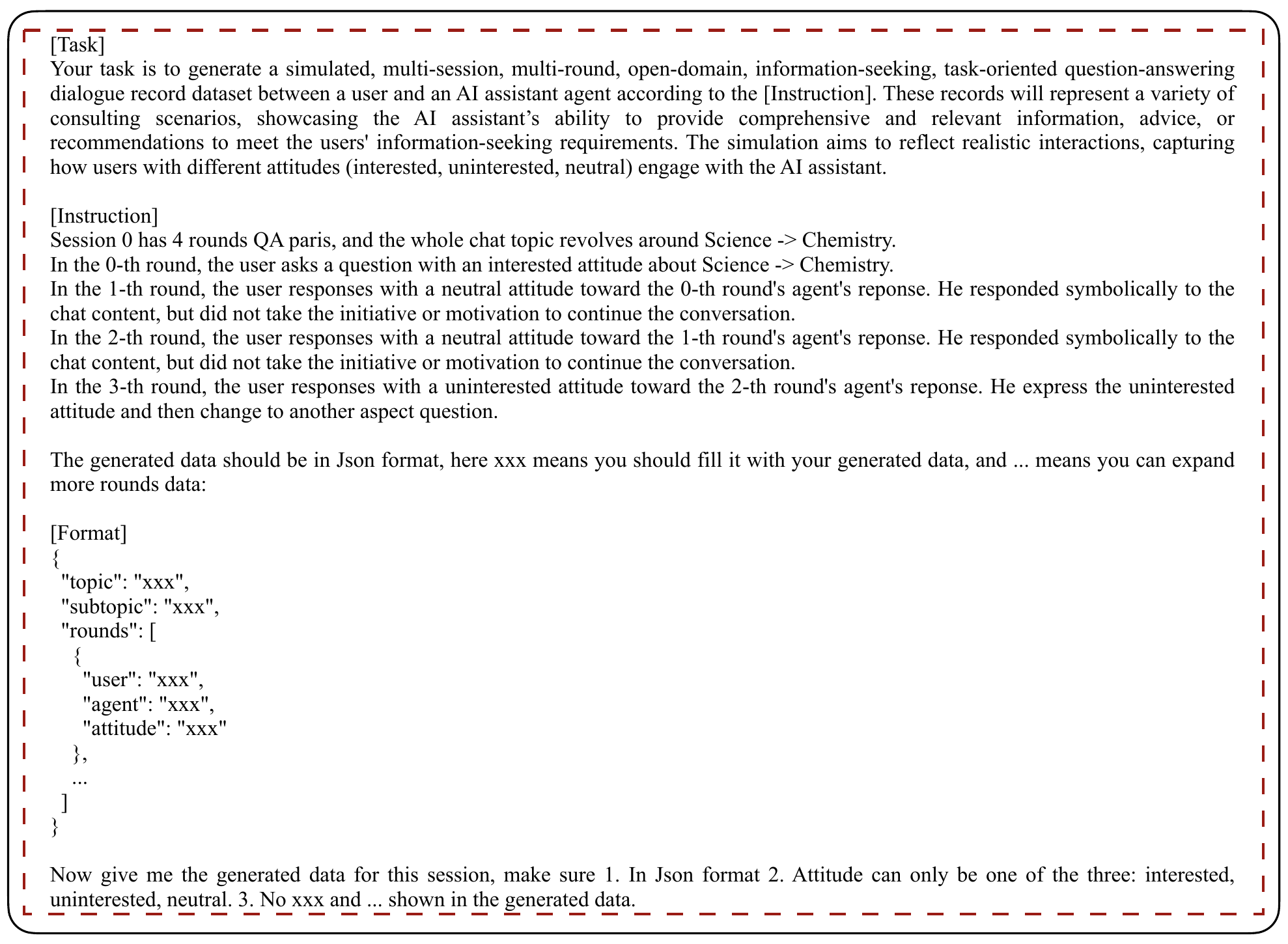}
    \vspace{-5mm}
   \caption{The prompt for the generation of simulated MSMTPInfo dataset. The instruction part of this prompt is generated by a stochastic process controlling the flow of the dialogue and the user's attitude.}
    \label{fig: Prompt_MSMTPInfo}
\end{figure*}

\begin{figure*}[h]
  \centering
  \includegraphics[width=5.0 in]{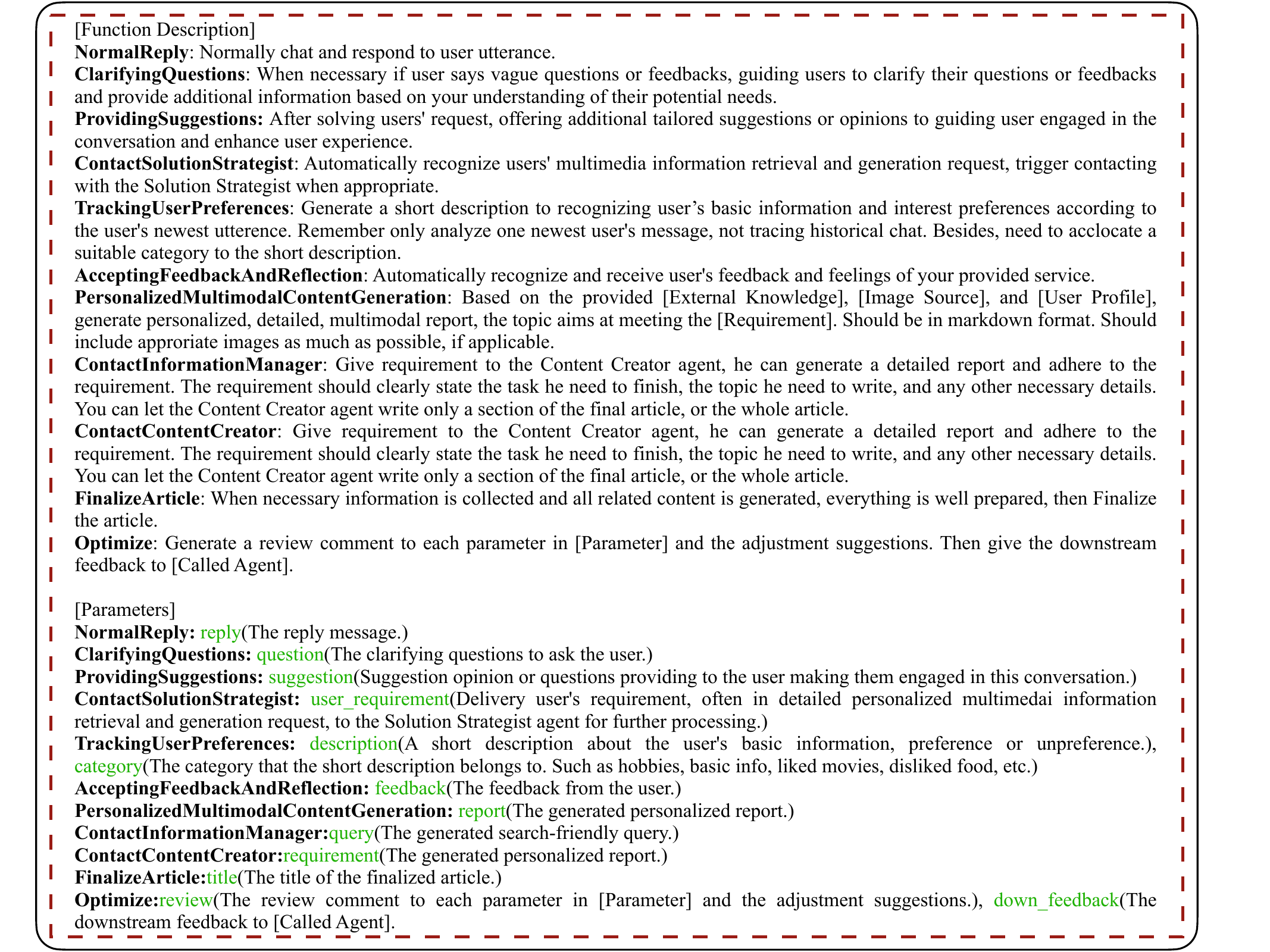}
  \vspace{-3mm}
  \caption{A overview of function prompts and parameter settings for function calls.}
  \label{fig: tool_prompt}
\end{figure*}

\section{Prompt}
\subsection{Synthetic MSMTPInfo Dataset Generation}
The MSMTPInfo dataset encompasses 13 main topics along with various subtopics. User response attitudes are categorized into three distinct types, each corresponding to different reactions, as detailed in Figure~\ref{fig: topic_attitude}. Building on this framework, we employed GPT-3.5 to simulate real user interactions with an AI assistant. We generated conversations session by session, with each session comprising a randomly determined number of dialogue turns centered around a main topic-subtopic configuration. The entirety of the session's dialogue is structured to revolve around the designated thematic setting. The prompt is illustrated in Figure~\ref{fig: Prompt_MSMTPInfo}.

\subsection{Function Call Design}
\label{sec: functionc_call_prompt}
In this section, we have systematically detailed all the design for function call prompts and parameter settings, as illustrated in Figure~\ref{fig: tool_prompt}. This figure comprehensively represents the prompts and parameters associated with each available function that can be invoked by agents.

\end{document}